\documentclass[draft]{agujournal2019}
\usepackage{url} 
\usepackage{lineno}
\usepackage[inline]{trackchanges} 
\usepackage{soul}
\usepackage{float}
\journalname{arXiv}

\begin{document}
\justifying

\title{Fiber-optic detection of snow avalanches using telecommunication infrastructure}


\authors{P. Edme\affil{1}, P. Paitz\affil{2}, F. Walter\affil{2}, A. Van Herwijnen\affil{3} \ and A. Fichtner\affil{1}}

\affiliation{1}{Institute of Geophysics, ETH Zurich, Zurich, Switzerland}
\affiliation{2}{Swiss Federal Institute for Forest, Snow and Landscape Research WSL, Birmensdorf, Switzerland}
\affiliation{3}{Swiss Federal Institute for Forest, Snow and Landscape Research WSL, Davos, Switzerland}

\correspondingauthor{Pascal Edme}{pascal.edme@erdw.ethz.ch}


\vspace{0.5cm}

\begin{abstract}
We demonstrate the detectability of snow avalanches using Distributed Acoustic Sensing (DAS) with existing fiber-optic telecommunication cables. For this, during winter 2021/2022, we interrogated a $\sim$10 km long cable closely following the avalanche prone Fl\"{u}elapass road in the Swiss Alps. In addition to other signals like traffic and earthquakes, the DAS data contain clear recordings of numerous snow avalanches, even though most of them do not reach the cable. Here we present two examples of snow avalanche recordings that could be verified photographically. Our results open new perspectives for cost-effective, near-real-time avalanche monitoring over long distances using pre-installed fiber-optic infrastructure.
\end{abstract}


\section{Introduction}

\noindent As a result of climate change and the related increase of extreme weather events, there is a growing hazard by mass movements to both population and critical infrastructure worldwide. Landslides, avalanches and flash floods frequently pose a significant risk to the population, with thousands of fatalities each year and billions of dollars in financial damage \cite{dilley2005natural, petley2012global, froude2018global, emberson2020new}. Early detection and mitigation require extensive, large-scale monitoring.

 \noindent Operational monitoring systems rely on camera and radar observations, tripwires, as well as infrasound and seismic sensors \cite{allstadt2018seismic, hurlimann2019debris}. The latter two have been shown to detect and identify mass movements \cite{schimmel2018automatic}, and recent developments in machine learning further improved their detection and warning capabilities for debris flows \cite{chmiel2021machine}. However, limitations in both resolution and spatial coverage remain for all currently available systems.

\noindent Emerging fiber-optic sensing techniques may help to overcome these issues. Distributed Acoustic Sensing (DAS), in particular, effectively transforms a standard telecommunication optical fiber into a distributed deformation sensor with a measurement point spacing as small as 25 cm. The DAS system consists of an Interrogation Unit (IU) that emits and receives laser pulses. Hence, a single instrument, together with the fiber, forms a distributed sensing antenna with thousands of measurements along a fiber of up to several tens of kilometers length. DAS makes installations and real-time monitoring with such a large number of measurement points logistically feasible and cost-effective \cite{hartog2017introduction,lindsey2021fiber}. 

\noindent The DAS instrument response has been studied over a wide range of frequencies \cite{lindsey2020broadband,paitz2021empirical}, and the robustness of current IUs has opened new opportunities in cryosphere research \cite{walter2020distributed,klaasen2021distributed,Hudson_2021,Fichtner_2022b,booth2020distributed,Fichtner_2023}. Specifically in the context of snow avalanches, \citeA{Paitz_2022} demonstrated that optical fibers can be used to measure ground deformation induced by avalanches. In their study, the avalanches propagated along the fiber at a dedicated test site, which is not representative of realistic monitoring scenarios.

\noindent Here we present results from interrogation of a telecommunication fiber along a mountain pass in Switzerland during winter 2021/22. The cable crosses several avalanche-prone sections of the pass road providing a commonly encountered situation that demands monitoring and warning solutions.

\section{Experimental setup}

\begin{figure}
\noindent\includegraphics[width=\textwidth]{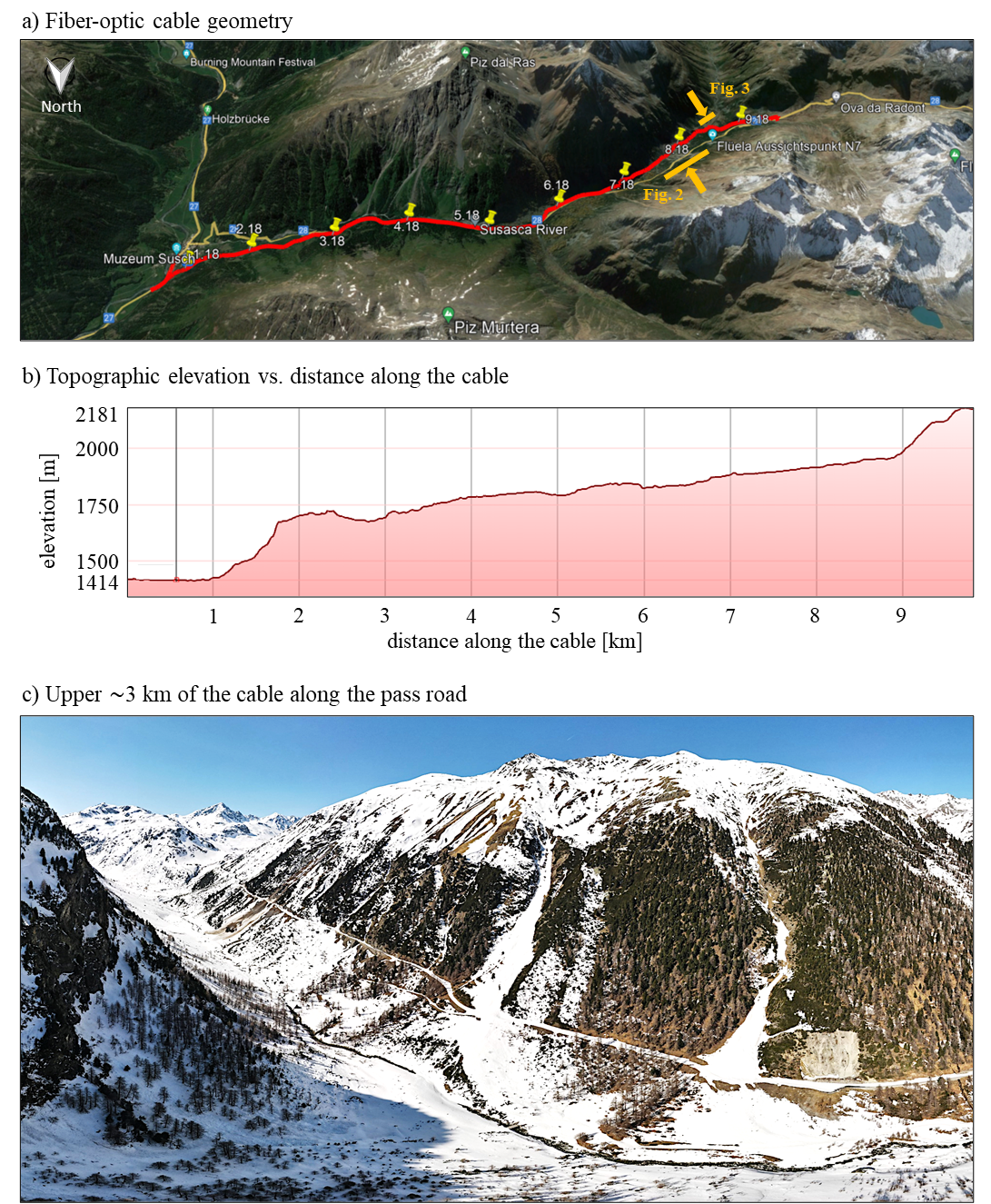}
\caption{Summary of the experimental setup. a) Geometry of the fiber-optic telecommunication cable starting in Susch and leading up to the Fl\"{u}elapass top. The cable following the pass road is marked in red. Yellow pins are distance markings in km. Thick orange lines indicate the locations of the avalanche examples shown in Figs. \ref{F:example1} and \ref{F:example2}, respectively. Source: Google Earth. b) Topographic elevation along the cable. Source: Google Earth. c) Drone image of the upper $\sim$3 km of the fiber-optic cable along the Fl\"{u}elapass road where most snow avalanches occurred. Picture credit: Lars Gebraad.}
\label{F:setup}
\end{figure}

\noindent From 23 December 2021 to 9 May 2022 we interrogated a $\sim$10 km long existing fiber-optic cable along the Fl\"{u}elapass, a high mountain pass road in the Swiss Alps, with a Silixa iDAS$^{TM}$ 2.0. Since the location is well-known to suffer from abundant snow avalanches, the road is closed during the avalanche season for about three months per year. The elevation of the cable ranges from 1414 to 2181 m above sea level. The fiber-optic cable geometry and the topography of the cable, as well as a photograph of the upper $\sim$3 km of the cable are visualized in Fig. \ref{F:setup}. 

\noindent We located the cable with numerous tap tests, leading to an estimated accuracy of the channel mapping of $\sim$20 m. The interrogator was located in the basement of a Swisscom fiber distribution hub in the village of Susch, and the first $\sim$6 km of the fiber closely follow the mountain pass road. For the upper $\sim$4 km, the cable follows the south side of the Susasca River in the valley, with a maximum distance of $\sim$350 m from the road. We recorded the raw DAS data with a sampling frequency of 100 Hz and 2 m channel spacing, resulting in $\sim$5000 channels.

\noindent To validate suspected avalanche records with ground truth observations, we performed several field visits and used a drone to collect photographic evidence, as shown, for example, in Fig. \ref{F:example1}a. In addition, we installed a camera near the end of the cable, viewing downvalley towards the north-facing slopes in the southern part of the valley (Fig. \ref{F:example2}a).

\section{Examples}

\begin{figure}[H]
\noindent\includegraphics[width=5.45 in]{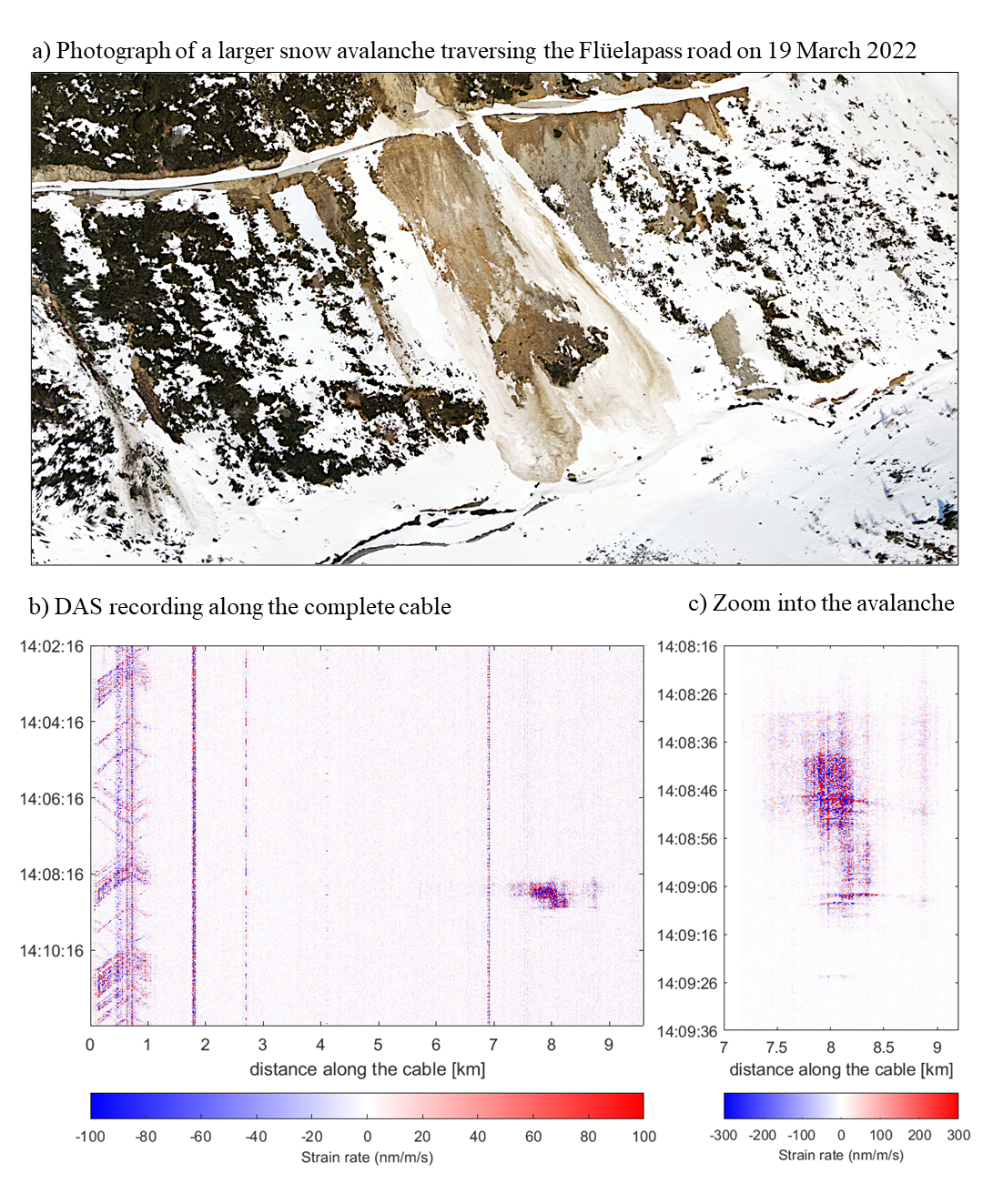}
\caption{Example of a larger snow avalanche that was clearly recorded along $\sim$1.5 km of the fiber-optic cable. a) Photograph of the snow avalanche traversing the Fl\"{u}elapass road and reaching the Susasca river in the valley (opposite the cable location). b) DAS recording along the complete $\sim$10 km long fiber-optic cable. Recordings at $<$1 km are mostly cars on the Cantonal road through the Engadin valley, which is not affected by winter closure of the Fl\"{u}elapass road. The snow avalanche recording is between $\sim$7.5 to 9 km distance. c) Zoom into the snow avalanche recording.}
\label{F:example1}
\end{figure}

\begin{figure}[H]
\noindent\includegraphics[width=\textwidth]{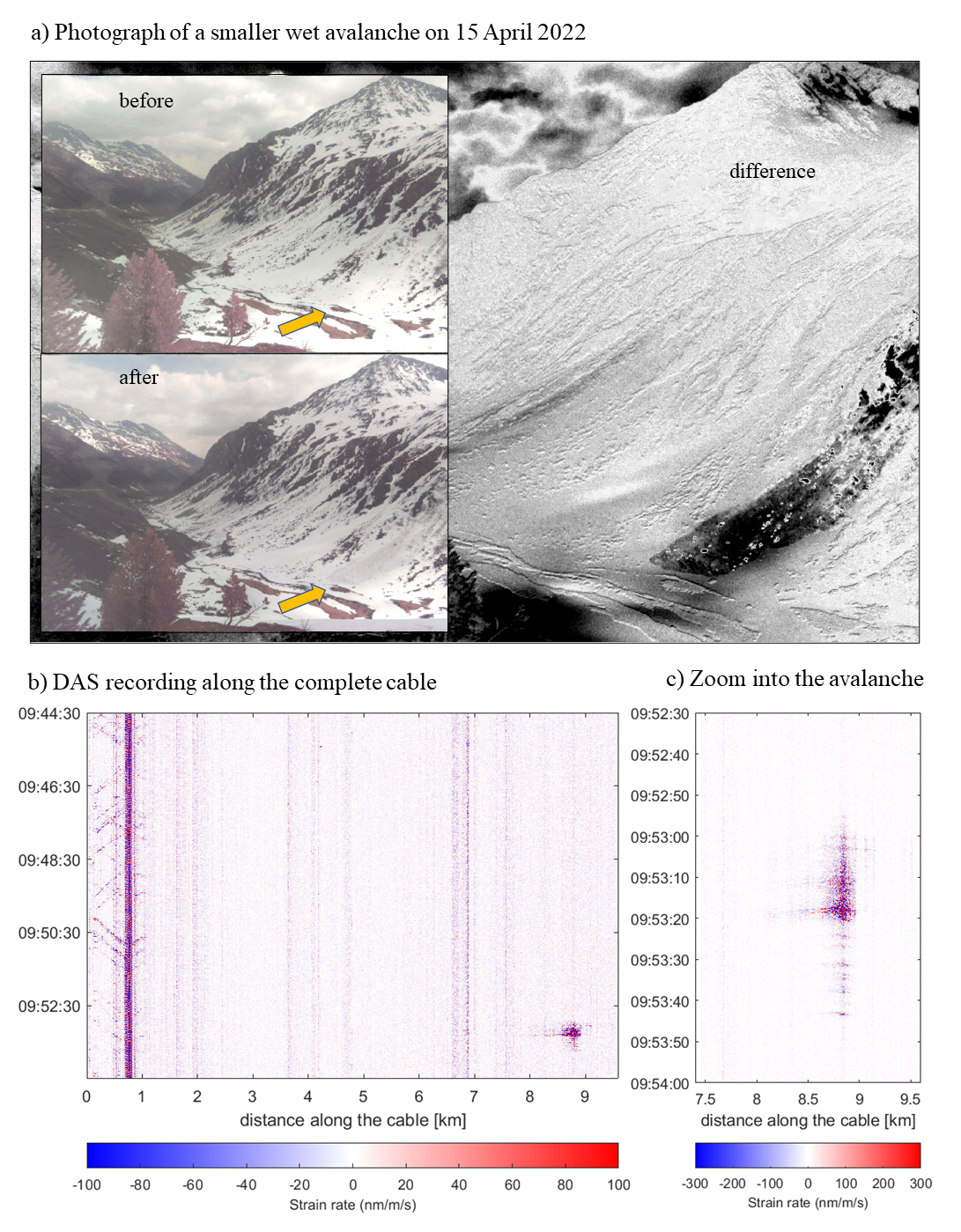}
\caption{Example of a smaller snow avalanche from 15 April 2022. a) Photographs taken before and after the occurrence of the snow avalanche (10min interval). Shown in gray scale is the difference (contrast boosted) between the two pictures. b) DAS recording of the snow avalanche around 9 km distance along the fiber-optic cable. }
\label{F:example2}
\end{figure}

\noindent During the experiment, we recorded a wide range of signals along the cable. They include signals from hikers and animals, earthquakes and numerous avalanche recordings, two examples of which are shown below. During the beginning and the end phase of the experiment, the mountain pass was open, and traffic noise is also amongst the recorded signals. Figures \ref{F:example1} and \ref{F:example2} show time windows during which avalanches were suspected. The background noise level is around 20 $\eta$m/m/s.  Spatially coherent signals at the beginning of the cable (distances $<$ 1.2 km) correspond to the vehicle traffic within the village of Susch, with velocities slower than 40 km/h and induced strain fluctuations not exceeding 1 $\mu$m/m/s.

\noindent With drone images we could confirm a large avalanche traversing the pass road at around 8 km distance along the fiber on 19 March 2022. The event is shown in Fig. \ref{F:example1} and can be observed over about 1.5 km distance along the fiber, even though its visible physical extent is $<$ 200 m. The snow mass stopped at the very bottom of the valley, on the north side of the river and therefore did not propagate over the cable situated slightly further up on the south side. The avalanche results in a quite complex signal, including seismic arrivals of apparent velocities faster than 1500 m/s.    

\noindent We confirmed another avalanche event on 15 April 2022 by taking the difference of two subsequent images from the camera installed at the end of the fiber, as shown in Fig. \ref{F:example2}. Despite happening on the south side of the river, this relatively small avalanche did not traverse the fiber but stopped just before it. The event is still visible over a few hundreds of meters in the DAS data, as a result of the seismic waves it generated. The induced strain fluctuation reached a level of  3 $\mu$m/m/s, slightly lower than the larger event of figure \ref{F:example1}.

\section{Discussion and conclusions}

\noindent We demonstrated that DAS with existing telecommunication cables can be used to detect snow avalanches, even of moderate size, and even though they do not reach the cable itself, as confirmed photographically. We believe that avalanche events can be discriminated from other signal like traffic, considering for example their fast apparent velocities and quite complex nature compared to the slow spatially consistent signal induced by vehicles.

\noindent These initial results shows that alpine natural hazard monitoring with fiber-optic is possible at least for snow avalanches and probably for other mass movements like rock falls. The DAS based cost effective solution opens new perspectives for near-real-time and early warning applications, in particular for critical infrastructure monitoring such as roads, railways, dams, and tunnels.

\noindent In many regions of interest, there is existing fiber-optic infrastructure that can be leveraged, thereby alleviating the need to install a large number of conventional instruments difficult to deploy and maintain. Furthermore, thanks to its high spatial and temporal sampling, DAS also provides information on traffic, which may assist in the estimation of potential damage caused by mass movements and can be crucial for first responders and rescue teams.


\acknowledgments
We gratefully acknowledge the support by Swisscom in the form of free access to the telecommunication cable along the Fl\"{u}elapass road. The drone footage was collected by Lars Gebraad.


\bibliography{biblio}

\end{document}